# Sinusoidal wave generating network based on adversarial learning and its application: synthesizing frog sounds for data augmentation

Sangwook Park, David K. Han, and Hanseok Ko

*Abstract*—Simulators that generate observations based on theoretical models can be important tools for development, prediction, and assessment of signal processing algorithms. In order to design these simulators, painstaking effort is required to construct mathematical models according to their application. Complex models are sometimes necessary to represent a variety of real phenomena. In contrast, obtaining synthetic observations from generative models developed from real observations often require much less effort. This paper proposes a generative model based on adversarial learning. Given that observations are typically signals composed of a linear combination of sinusoidal waves and random noises, sinusoidal wave generating networks are first designed based on an adversarial network. Audio waveform generation can then be performed using the proposed network. Several approaches to designing the objective function of the proposed network using adversarial learning are investigated experimentally. In addition, amphibian sound classification is performed using a convolutional neural network trained with real and synthetic sounds. Both qualitative and quantitative results show that the proposed generative model makes realistic signals and is very helpful for data augmentation and data analysis.

## I. Introduction

PHYSICAL model based simulators are important tools for signal processing algorithms. In audio signal processing, RoomSim has been widely used to obtain spatial audio data across several research areas, such as sound source localization, separation, and de-reverberation [1]–[6]. For developing noise robust recognition systems, AddNoise is used to compose noisy databases [7], [8]. For other purposes, simulators based on regression models can be used to predict benefits, costs, and any feasibility in the future. In radar signal processing, simulators incorporating several models of propagation, reflection, and antenna responses were applied for performance assessments under various conditions [9], [10]. Kim, et al. [11] have improved speech recognition performance based on Long-Short Time Memory (LSTM) by using a room simulator. As these examples show, simulators are very helpful tools across the field of signal processing. However, designing a simulator requires painstaking efforts to make models of each affective element. Also, complicated models are usually required to represent the real phenomena occurring in a variety of environmental conditions.

On the other hand, recently generative models based on deep networks have been proposed for synthesizing observations. Since generative models can be constructed by directly training with real observations, the painstaking effort required to establish theoretical models during simulator design would be avoided. Donahue, et al. [12] used this room simulator to demonstrate their speech enhancement method in additive and reverberant noisy conditions. For classification tasks using deep networks, the idea of using generative models to augment data has begun to attract the attentions of researchers, because deep networks require significant quantities of training data. Traditionally, probabilistic approaches based on Maximum Likelihoods (MLs) are considered for making generative models. However, to make a ML based generative model requires a large number of training data depends on a synthesized data dimension known as the *Curse of Dimensionality* [13]. Alternatively, Goodfellow, et al. [14] proposed a Generative Adversarial Network (GAN) that could be considered as another way to make a generative model, and demonstrated the feasibility of generating images from a random vector. Recently, several works considered augmenting training data in image classification using synthesized images [15]–[17], and in acoustic scene classification using synthesized audio features [18].

Given that the GAN based generative model produces training datasets, this paper tries to expand the approach to arbitrary signal generation using GAN. To achieve this purpose, this paper proposes a sinusoidal waveform generating network first because signals are composed by a linear combination of sinusoidal waveforms. Secondly, the proposed network is tested using the examples of both sinusoidal waveform generation and amphibian call sound generation. Additionally, amphibian sound classification is performed using Convolutional Neural Networks (CNNs) trained by both/either real and/or the synthesized amphibian sounds. Our main contributions are as follows: 1) development of the link connecting two fields, signal processing and machine learning; 2) development of a GAN based generative model for generating sinusoidal waveforms; 3) demonstration of the proposed network by synthesis of amphibian sounds; and 4) validation of the effectiveness of the data augmentation using synthesized data from the proposed network.

The rest of this paper is organized as follows. The next section investigates related works on adversarial learning using deep networks and generative models for waveform generation.

Section III describes the proposed network for generating sinusoidal waveforms using adversarial learning. In Section IV, several adversarial learning approaches are investigated for training the proposed network, and amphibian sound generation using the proposed method is tried. In Section V, the efficiency and effectiveness of data augmentation using the proposed method are demonstrated through amphibian sound classification. Conclusions are given in the final section

## II. RELATED WORKS

### A. Adversarial learning

Goodfellow et al. [14] proposed an adversarial modeling of a generator $G(z;\theta_G)$ and a discriminator $D(x;\theta_D)$. The generator $G$ maps a random initialized vector $z$ into a data space while the discriminator $D$ represents the probability that $x$ came from the real data space. The training progresses by optimizing the objective function, $f_{GAN}(\theta_G,\theta_D)=E_{Pdata}(x)[log(D(x;\theta_D)]+E_{Pz(z)}[log(1-D(G(z;\theta_G);\theta_D)]$ where $E_{Pdata}[.]$ and $E_{Pz(z)}[.]$ means an expectation operator to probability density function of real data and random vector, respectively. The parameters of the generator $\theta_G$ are trained toward minimizing the function while the parameters of the discriminator $\theta_D$ wants to maximize the function. In practice, $\theta_G$ and $\theta_D$ are alternatively and repetitively trained under the condition that the other is fixed [14] as follows.

$$\theta_D^n = \arg\max_{\theta_D} f_{GAN}(\theta_G^{n-1}, \theta_D) \\ \theta_G^n = \arg\min_{\theta_G} f_{GAN}(\theta_G, \theta_D^n),$$
(1)

where $n$ is the number of iterations. Following their work, *Conditional GAN* (CGAN) was proposed by specifying additional conditions such as class labels [19], and deep convolutional GAN was developed with unsupervised representational learning using CNN [20]. For generating high quality images based on CGAN, other methods have been proposed [21]. Reed et al. [22] proposed a new method that generates images from scene descriptive captions.

As mentioned previously, these approaches suffer from several issues such as a lack of optimization, unbalanced learning between the two networks, and mode collapse due to the alternating training [23]–[25].

### B. Approaches for stable learning

To address the lack of optimization stemming from the coupled cost functions, the objective function itself is modified for successful optimization in other methods. Mao et al. [26] proposed Least Squares GAN (LSGAN) that adopted the least square error as the cost $f_{LSGAN}(\theta_G; \theta_D) = E_{Pdata(x)}[(D(x; \theta_D) - 1)^2] + E_{Pz(z)}[(D(G(z; \theta_G); \theta_D))^2]$. They have been reported that their proposed least squares loss function forces the synthetic data toward the decision boundary resulting the synthetic data being in closer proximity to real data and the LSGAN results in more stable than a vanilla GAN. However, the least square error may not be the most appropriate measure for determining the proximity of synthetic data to real data.

Arjovsky et al. [23] proposed *Wasserstein GAN* (WGAN) for stable learning. In WGAN, Earth-Mover (EM), i.e. Wasserstein distance was used to design an objective function instead of alternatives such as Kullback-Leibler (KL) divergence or Jensen-Shannon (JS) divergence. According to the paper, KL or JS divergence is inappropriate as an objective function for GAN due to their discontinuities although they are well known as a metric to measure the differences between two distributions. On the other hand, the Wasserstein distance is continuous and differentiable in most of the parameter domain if the discriminator satisfies the Lipschitz condition. The function applied in WGAN is defined as $f_{WGAN}(\theta_G; \theta_D) = E_{Pdata(x)}[D(x; \theta_D)] - E_{Pz(z)}[D(G(z; \theta_G); \theta_D)]$. Both $\theta_G$ and $\theta_D$ are updated toward maximizing the function. In order to satisfy the Lipschitz condition, $\theta_D$ has to be clipped on the interval [-c, c]. But, it is hard to optimize the parameter $c$. Instead of the clipping method, Gulrajani, et al. [27] proposed a new way to satisfy the condition that adds the gradient penalty to the function as

$$f_{WGAN\_GP} = f_{WGAN} - \lambda_{GP} E_{\hat{x}}[(\|\nabla_{\hat{x}} D(\hat{x})\|_2 - 1)^2],$$
(2)

where $\hat{x}$ is an internally dividing data between the real and synthetic data from the generator, and $\lambda_{GP}$ is a constant parameter. By using the gradient penalty, the Lipschitz condition can be satisfied without weight clipping. However, the parameter $\lambda_{GP}$ is too sensitive to train adversarial networks. In order to resolve this issue, Petzka, et al. [28] recently proposed a modified gradient penalty method as

$$f_{WGAN\_LP} = f_{WGAN} - \lambda_{LP} E_{\hat{x}}[(\max\{0, \|\nabla_{\hat{x}} D(\hat{x})\| - 1\})^2].$$
(3)

In other research, Arjovsky and Bottou [29] investigated the reasons for unstable learning in GAN and proposed a learning technique for stable learning. The technique adds a little perturbation to both the original and the synthetic data to resolve the problem of dimensional misspecification when a discriminator is trained.

### C. Approaches for generating audio waveforms

In this section, research on conventional audio generation by generative models is summarized since the 1D sequential data format of audio waveforms more closely resembles a general signal than an image. As mentioned previously, a significant amount of data is required to train ML based generative models in a high-dimensional space. For example, at least 240k of data is required to build a generative model in 48k dimensional space for generating 3 second audio sequences sampled at 16 kHz [30]. In the case of music generation, the synthesized dimension can be reduced by using symbolic data, i.e. MIDI, instead of raw audio waveforms. Because a single MIDI note can last a second and be represented by note, velocity, and duration, the dimension of audio sequence can be dramatically reduced. MelodyRNN which is a well-known approach for symbolic-domain music generation consists of three Recurrent Neural Networks (RNNs), a RNN to learn longer-terms, a lookback RNN, and an attention RNN [31]. As for other approaches, a hierarchical RNN structure has been proposed to generate melody, chords, and drums in Song from PI [32], and DeepBach generates *J. S. Bach* style music depending on user constraints such as rhythm, notes, parts, chords, and cadences [33]. C-RNN-GAN is known as a GAN based model with both the generator and the discriminator constructed by LSTM [34]. MidiNet is also a GAN based model with convolutional structures [35]. In both of these GAN based symbolic domain music generation models, $f_{GAN}$, is used to train adversarial networks.

For more general audio generation, such as speech generation, the dimension of the waveform composed by sequential samples can be reduced by applying the autoregression model. Under this approach, a new sample is estimated from several previous samples. To give a representative example, WaveNet can be reduced trainable parameters by applying dilated convolution to previous samples to obtain a large-sized receptive field [36]. Since the network would be trained without information of signal duration, however, the network will generate noise that has a large portion in large-sized receptive field in case of synthesizing a short-time audio waveform. As a GAN based generative model, WaveGAN is designed according to a 1D convolutional structure in terms of its generator and discriminator [37], which are trained by optimizing $f_{WGAN\_GP}$.

### III. SIGNAL GENERATION USING GAN

To develop a general signal generation model, technical approaches from signal model to GAN training are described in this section. The signal model is firstly reconsidered to find an approach for reducing a dimension of the synthesized signal space, and network architecture is then suggested for GAN based signal generation. A search for successive training of the proposed networks follows.

#### A. Signal model in the aspect of neural network

In many signal processing methods, signals are typically interpreted in the frequency domain by applying Fourier transform given that an arbitrary signal is represented as a linear combination with frequency bases. Note that a frame composed of consecutive samples within a short-time can be considered as quasi-stationary even if a signal is generally considered nonstationary. In the real environment, a frame sample in observation, $x_n$, can be represented by adding noise to signal under an additive noise condition as in

$$x_n = \frac{1}{K}\sum_{k=0}^{K-1} w_k exp\left(\frac{j2\pi nk}{K}\right) + d_n, \quad 0 \leq n \leq K-1. \quad (4)$$

where $K$ is the length of a frame, and $w_k$ and $d_n$ means the weight and noise, respectively. Typically, the noise term is considered as an independent and identically distributed random variable along with $N(0;\sigma^2)$. Note that both weights and bases are typically considered in the complex domain.

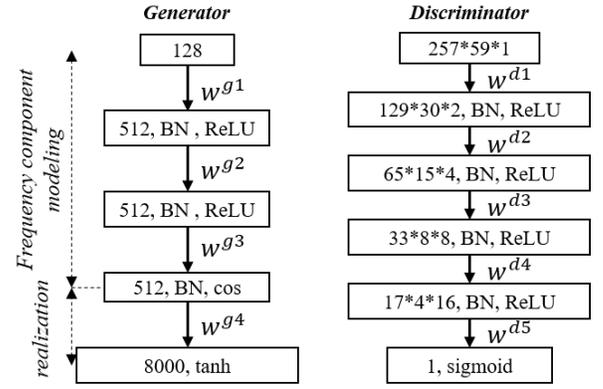

Fig. 1. Network architectures of generator and discriminator for signal generation; the weight, $w^{g4}$, for realization satisfies the weight condition as $w^{g4}(i, j) = w^{g4}(i+256, j)$, for $1\leq i \leq 256, 1\leq j \leq 8000$.

For considering real signal generation based on the sample model, a signal part can be derived as

$$s_n = \frac{1}{K}\sum_{k=0}^{K-1} w_k exp\left(\frac{j2\pi nk}{K}\right) \quad (5)$$
$$= \frac{1}{K}\left(w_0 + w_{K/2} + \sum_{k=1}^{K/2-1}\left\{w_k exp\left(\frac{j2\pi nk}{K}\right) + w_{K-k} exp\left(-\frac{j2\pi nk}{K}\right)\right\}\right).$$

As shown in (5), half of the complex bases are conjugate bases of the other half of the bases. If both weights, $w_0$ and $w_{K/2}$, are real and $w_k = w^*_{K-k}$ for $1 \leq k \leq K/2$-1, then the real signal can be represented as

$$s_n = \frac{1}{K}\left(w_0 + w_{K/2} + 2\sum_{k=1}^{K/2-1} re(w_k)cos\left(\frac{2\pi nk}{K}\right)\right). \quad (6)$$

where $re(\cdot)$ is an operator for returning the real part of a complex number. From the neural network perspective, this can be implemented by considering that the cosine function is applied as an activation function in the last hidden layer, $re(w_k)$ is the weight connecting between the last hidden layer and output layer, and the remainders, $w_0$ and $w_{K/2}$ are a bias in the output layer. This part is called by *realization* that means real audio sample would be realized from frequency component.

#### B. Network architectures for generator and discriminator

Based on the investigation of signal models, Multi-Layer Perceptron (MLP) is applied to the generator in the proposed adversarial networks, as depicted in Fig. 1. The generator can be divided into two parts, frequency component modeling and realization. The frequency components for representing a signal are modeled through three fully connected layers with Batch Normalization (BN) [38]. Since the frequency is considered as non-negative, Rectified Linear Unit (ReLU) activation is applied to the first two layers [39]. In the second part, the cosine function is applied in the third layer as mentioned previously. For implementing signal generation based on (6), the weights connecting to the output layer are set to real and half of the weights are the same as the remainders to satisfy the weight constraints. A hyperbolic tangent is used as an activation function in the output layer because the output nodes mean samples in the waveform. In this architecture, since the third layer nodes are connected to all samples in a waveform, it can be argued that the frequency bases are shared by all the frames. And, the weight that connects the uncorrelated sample and the frequency basis would diminish as the network is trained. In this paper, the sampling rate of the synthesized waveform is set to 16 $kHz$, and the number of nodes in the intermediate layers is determined by considering the length of a frame to be 32 ms (512 points). As shown in Fig. 1, 0.5 second audio waveform would be synthesized from random initialized 128 dimensional vector by this generator after training.

CNN based discriminator composed of 4-convolutional layers and 1-fully connected layer is applied to the proposed network. Both the real waveform and the waveform synthesized by generator are transformed to a spectrogram using the Short Time Fourier Transform (STFT). Under this configuration, the frame and sliding length is set to 32ms and 8ms, respectively. At the end of each convolutional layer, both the width and height of the mid-layer spectrum are reduced to as little as half of the previous mid-layer spectrum by applying max-pooling. The convolution filter size is set to 3x3x$N_i$ where $N_i$ is the third dimension of the previous layer. BN and ReLU are then applied. After the fourth convolution and pooling, the 3-dimensional results are flattened to a 1-dimensional vector by constructing a fully-connected layer. The last weight denoted as $w^{d5}$ in Fig. 1 is the weight connecting the flattened layer to the output layer.

### C. Adversarial learning of both networks

An objective function for training the proposed network, $f_{prop.}$ is defined as

$$f_{prop.} = f_{obj} - \lambda \sum_{i,j} \left| w^{g4}_{i,j} \right|. \tag{7}$$

The weights $w^{g4}_{i,j}$ that connect to the output layer in the generator are applied to regularize with coefficient $\lambda$. To understand this regularization strategy further, note that the training data includes noise uncorrelated to signals in real applications. Moreover, unfortunately, the network cannot distinguish between signal and noise in observations when being trained with the training data. Thus, if the number of trainable parameters becomes sufficient to model both signal and noise, then the generator makes very noisy waveforms. This is a kind of overfitting occurring in the generator. To resolve this issue, regularization is applied to the proposed objective function according to the assumption that a signal is typically composed of a finite number of frequency bases while noise can be represented by combining an infinite number of frequency bases. In order to investigate the issue of unstable learning depending on the objective functions $f_{obj}$, several functions such as $f_{GAN}$, $f_{LSGAN}$, and $f_{WGAN}$ are considered in experiments, as well as $f_{WGAN\_GP}$ and $f_{WGAN\_LP}$.

The training networks use the following settings: all trainable parameters are initialized by the normal distribution featuring a zero mean and a 0.01 standard deviation; an initial random vector $z$ is also initialized by the normal distribution featuring a zero mean and a 0.01 standard deviation; the learning rate is set to 1.0e-6; the regularization coefficient, however, depends on the objective functions; and the batch size is 32. The python implementation can be found on https://github.com/tkddnr7671/SinusoidalGAN.

### IV. EXPERIMENT I. SIGNAL GENERATION

In this section, the reproducibility and productivity of the proposed network are respectively demonstrated by synthesizing a mathematically produced sinusoidal waveform and an amphibian call sound collected in a real environment. Note that amphibian sounds are chosen as the application for this paper because they are distinguishable despite their short durations, and they have plenty of frequency components to represent themselves. The synthesized waveform produced by the proposed method is graphically compared to the original target signal, either the sinusoidal waveform or the amphibian sound, in form of a spectrogram with a time-domain waveform. Also, the inception score [40] of the synthesized amphibian sounds is considered for quantitative comparison to other prominent methods.

### A. Demonstration of reproducibility

*1) Database*

In this experiment, the database for training the proposed network was composed of target waveforms mathematically generated as

$$x_n = \alpha \sum_{c=1}^{C} \cos\left( \frac{2\pi f_c (1+\Delta f_c) n}{f_s} + \Delta \phi_c \right) + d_n,\ 1 \le n \le \frac{f_s}{2}. \tag{8}$$

where $C$ is the number of frequency components whose frequency is set to $f_c$ and $f_s$ is the sampling rate. The deltas denoted in front of $f_c$ and $\phi$ are the variations in frequency and phase, respectively, to represent any distortions. The variations and additive noise $d_n$ are represented by a random variable along to normal distribution. The constant parameter $\alpha$ is determined depending on the Signal to Noise Ratio (SNR).

In training phase, the Blackman-Harris window was applied to the signal to eliminate discontinuity at both side-ends of the signal. The sampling rate and length of the target waveforms were set to 16 $kHz$ and 0.5 seconds, respectively. Many experiments were performed using the target model and varying the number of frequency components and SNR. In the following subsections, the results in waveforms composed by mono- and double-frequency components are respectively represented because it would be possible to make more complicated waveform based on superposition if the network can generate simple waveforms like that. Also, more complicated case would be demonstrated by amphibian call sounds in the next section.

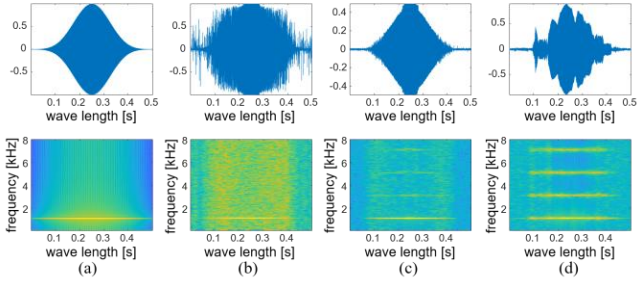

Fig. 2. Target signal whose frequency is 1.0 kHz under clean conditions (99dB) and synthetic signals by adversarial networks whose discriminator is composed of four fully-connected layers; (a) target signal, (b) GAN λ=1.0e-4, (c) LSGAN λ=1.0e-6, (d) WGAN λ=2.5e-6 & *c*=0.0015.

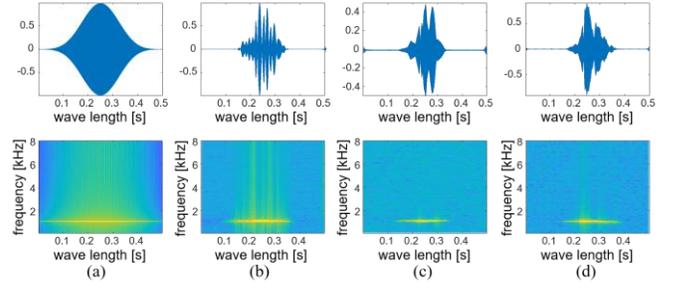

Fig. 3. Target signal whose frequency is 1.0 kHz under clean conditions (99dB) and synthetic signals by the proposed adversarial networks whose discriminator design is based on CNN; (a) target signal, (b) GAN λ=1.0e-4, (c) LSGAN λ=1.0e-4, (d) WGAN λ=1.0e-6 & *c*=0.005.

*2) Synthesizing mono-frequency sinusoidal waveform*

In adversarial learning, the target signal used for training the discriminator affects the waveforms generated by the generator. As such, the generator tries to mimic the target signal to deceive the discriminator, which tries to distinguish between the target and synthetic signal. Therefore, the generator will produce a realistic synthetic signal if the discriminator is sensitive at distinguishing between target signal and synthetic signal. This section tries to explain this investigation [16].

First of all, the mono-frequency target signal depicted in Fig. 2 (a) is produced based on (8) with the frequency set to 1.0 *kHz*, and α determined by satisfying the 99dB SNR condition. Instead of the CNN based discriminator described in Fig. 1A, a discriminator designed with four fully-connected layers is applied to adversarial learning. In this case, a discrete time-domain waveform is inputted directly to the discriminator. Synthesized signals of 1.0 *kHz* frequency are depicted in Fig. 2 (b) ~ (d) according to the objective function used. In the case of GAN, the synthesized signal seems like a signal distorted by noise even though the target signal is produced in clean conditions. The synthetic signals produced by the other approaches that were developed for stable learning, are cleaner than the result of GAN but exhibit unexpected harmonics. If $f_s=\delta f_c$, where δ is an integer, the target signal denoted by $x_n=\alpha\cos(2\pi n(1+\Delta f_c)/\delta+\Delta\phi)+d_n$ has no relation with the frequency $f_c$ anymore. In this configuration, the target signal can be approximated by $x_n=\alpha\cos(2\pi n/16)$, since a clean target was considered with δ=16. However, this can be denoted by $x^o_n=\alpha\cos(2\pi n/16)+\beta\cos(2\pi n3/16)+\beta\cos(2\pi n5/16)+\varepsilon\cos(2\pi n7/16)$ for odd samples and $x^e_n=\alpha\cos(2\pi n/16)+\varepsilon\cos(2\pi n7/16)$ for even samples if α > ε, because $\cos(2\pi n3/16)=(-1)^n\cos(2\pi n5/16)$ and $\cos(2\pi n/16)=(-1)^n\cos(2\pi n7/16)$. From the reason, training networks would be stopped when the generator can synthesize the waveform that shows unexpected harmonics as shown in Fig. 2 since the discriminator could not recognize these harmonics in time domain.

On the other hand, the spectrogram transformed from the discrete time-domain waveform is inputted into the CNN based discriminator depicted in Fig. 1. Since the harmonics are easily recognized in the spectrogram, the synthetic signals created by the proposed network have no unexpected harmonics and are clearer than the previous cases as shown in Fig. 3. In previous research, also CNN based networks have demonstrated the production of high quality synthetic data when using several different methods of generating images or audios [20], [36], [37], [41]. These results show that the CNN based discriminator is better than a fully-connected network for producing an arbitrary signal. Note that the regularization coefficients are experimentally determined depending on the type of objective function because the ranges of the objective functions differ. Also, in case of LSGAN, different reguarlization coefficients are required depending on network architecture as shown in Fig. 2 and Fig. 3.

*3) Synthesizing double-frequency sinusoidal waveform*

This subsection shows that, given that the arbitrary signal is represented as a linear combination with frequency bases, it generates a double-frequency sinusoidal waveform. A simple sinusoidal waveform having two frequencies, 1.0 kHz and 1.5 kHz, is used as the target signal. Fig. 4 shows the target signals and the synthetic signals formed using the proposed network, depending on the objective function. Even if the duration of the synthetic signal is shorter than the target duration, the multi-frequencies of the target signal are apparent in all the synthetic signals. However, the synthesized signal based on LSGAN shows approximately 1.25 kHz and 1.5 kHz frequencies that differ from the target signal. Except that, the synthetic signals have two harmonics at about 1.0 kHz and 1.5 kHz.

In multi-frequency sinusoidal waveform generation, the convolutional filter size in the discriminator intuitively affects a frequency resolution of the generator because it is hard to recognize a difference between two frequencies that are really close to each. However, this affect is also related to the number of frequency bins. For example, in the proposed discriminator the receptive field along the frequency domain is about 93.75 Hz when the number of frequency bins is set to 256. If the number of frequency bins is increased to 512, the receptive field along the frequency domain is reduced to about 46.87 Hz even if the filter size in the discriminator is fixed. Thus, it can be a method to acquire a generator having a high resolution in frequency domain to increase the number of frequency bins in STFT procedure. Obviously, it requires a modification of the input configuration in the discriminator.

*4) Effect on adversarial learning by noise*

In order to investigate the effect of noise on adversarial learning, the proposed network is trained using noisy signals of differing SNR. Note that the target signals produced in above the 20 dB SNR condition are considered. As an example, a target signal and the synthesized signals are shown in Fig. 5. In case of GAN, the noise is shown to affect the learning. Their multi-frequencies are indistinguishable to each other even if the regularization coefficient has to be experimentally adjusted. As mentioned previously, regularization is used to resolve overfitting problems arising in the generator, which can confuse noise for the signal when noisy data is used in adversarial learning. In cases of WGAN_GP and WGAN_LP, it is hard to prevent diminishing waveforms by adjusting both the regularization coefficient and the parameters $\lambda_{GP}$ and $\lambda_{LP}$ for each method.

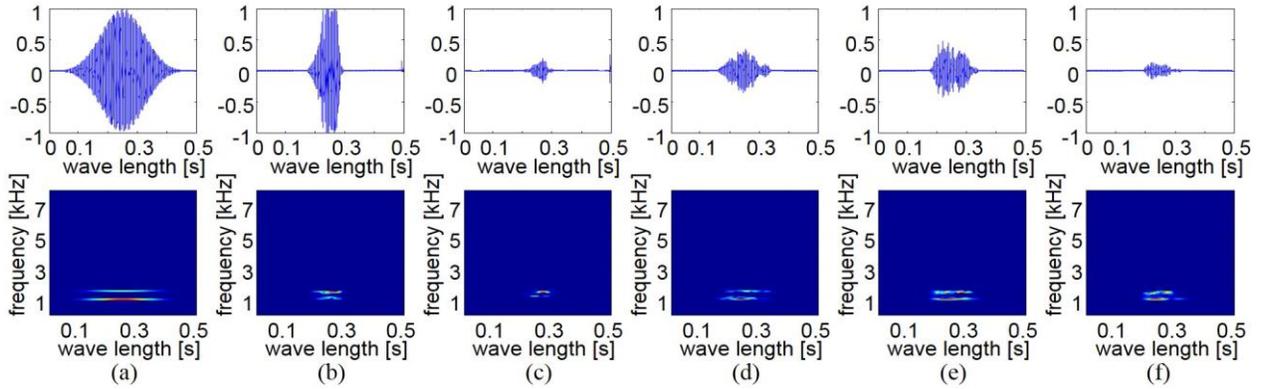

Fig. 4. Target signal composed of multi-frequencies 1.0 kHz and 1.5kHz under a clean condition (99dB) and synthetic signals by the proposed networks; (a) target signal, (b) GAN λ=1.0e-5, (c) LSGAN λ=0.5e-6, (d) WGAN λ=1.0e-6 & *c*=0.005, (e) WGAN_GP λ=1.0e-6 & $\lambda_{GP}$=20, (f) WGAN_LP λ=1.0e-6 & $\lambda_{LP}$=10.

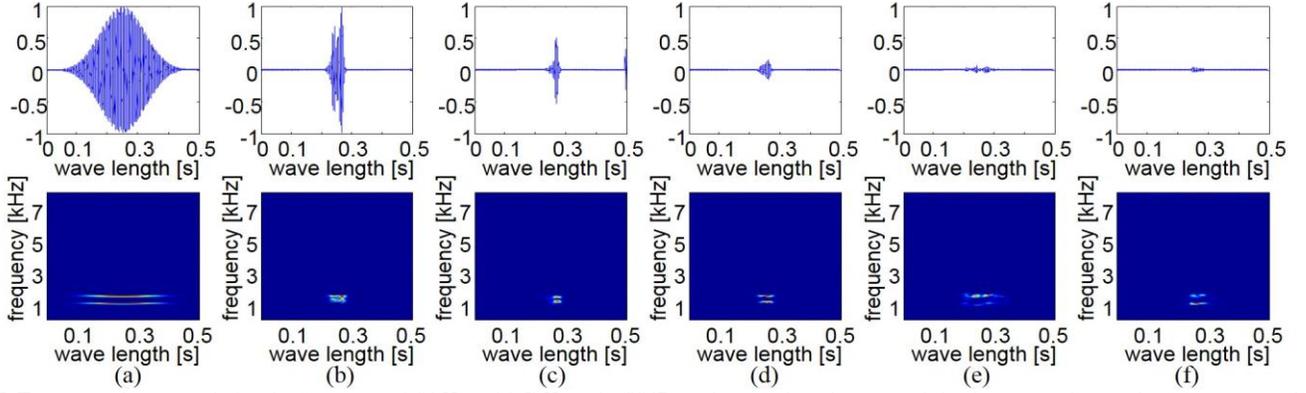

Fig. 5. Target signal composed of multi-frequencies 1.0 kHz and 1.5kHz under 20dB condition and synthetic signals by the proposed networks; (a) target signal, (b) GAN λ=1.0e-6, (c) LSGAN λ=0.5e-6, (d) WGAN λ=1.0e-6 & *c*=0.005, (e) WGAN_GP λ=1.0e-6 & $\lambda_{GP}$=10, (f) WGAN_LP λ=1.0e-6 & $\lambda_{LP}$=15.

On the other hand, the synthesized signal based on WGAN shows distinct multi-frequencies with a 0.005 weight clipping constraint. Although WGAN_GP and WGAN_LP were developed to resolve the problem of optimizing parameter *c* to satisfy the Lipschitz condition in WGAN, optimizing parameter $\lambda_{GP}$ or $\lambda_{LP}$ is also difficult in this configuration. For this reason, the objective function for training the proposed network is composed by the $f_{WGAN}$ with the weight clipping strategy where *c*=0.005 and regularization. This approach is also applied to the next experiment; amphibian call sound generation.

### B. Demonstration of productivity

#### 1) Database

For this experiment, amphibian call sounds collected in the natural habitats of each species were considered as the target waveform. The list of species can be found in TABLE I. After modifying the target audios to 16 kHz and mono-channel, they were divided into segments whose length was 0.5 seconds by applying the endpoint detection method [42]. As mentioned previously, if the length of a segment was less than 0.5 seconds, a random initialization featuring a zero-mean and 0.01 standard deviation was applied for network training instead of zero-padding.

#### 2) Graphical comparison

The proposed method which is trained using $f_{WGAN}$ with the regularization coefficient set to 2.5e-6 for an objective function in (7) is separately trained for each target species with common fixed parameters such as network architecture, batch size, and the random vector dimension of 128, as depicted in Fig. 1. As training data, 150 randomly selected segments were used per class. For three species marked as SuwFrog, RedFrog, and NarFrog, real and synthesized waveforms are shown in Fig. 6 with their spectrogram. (note that the first in Fig. 6 is the real data used in training the networks while the rest are synthetic waveforms). As shown in the figure, the synthesized waveforms are very similar to real ones except for their scale and amplitude shape.

TABLE I
THE LIST OF SCIENTIFIC NAMES FOR TARGET SPECIES AND ITS ABBREVIATION

| Scientific Name | Abbreviation | The number of data |
|---|---|---|
| Hyla suweonensis | SuwFrog | 5,303 |
| Hyla japonica | GreFrog | 1,664 |
| Kaloula borealis | NarFrog | 2,241 |
| Rana dybowskii | BroFrog | 676 |
| Bombina orientalis | RedFrog | 166 |

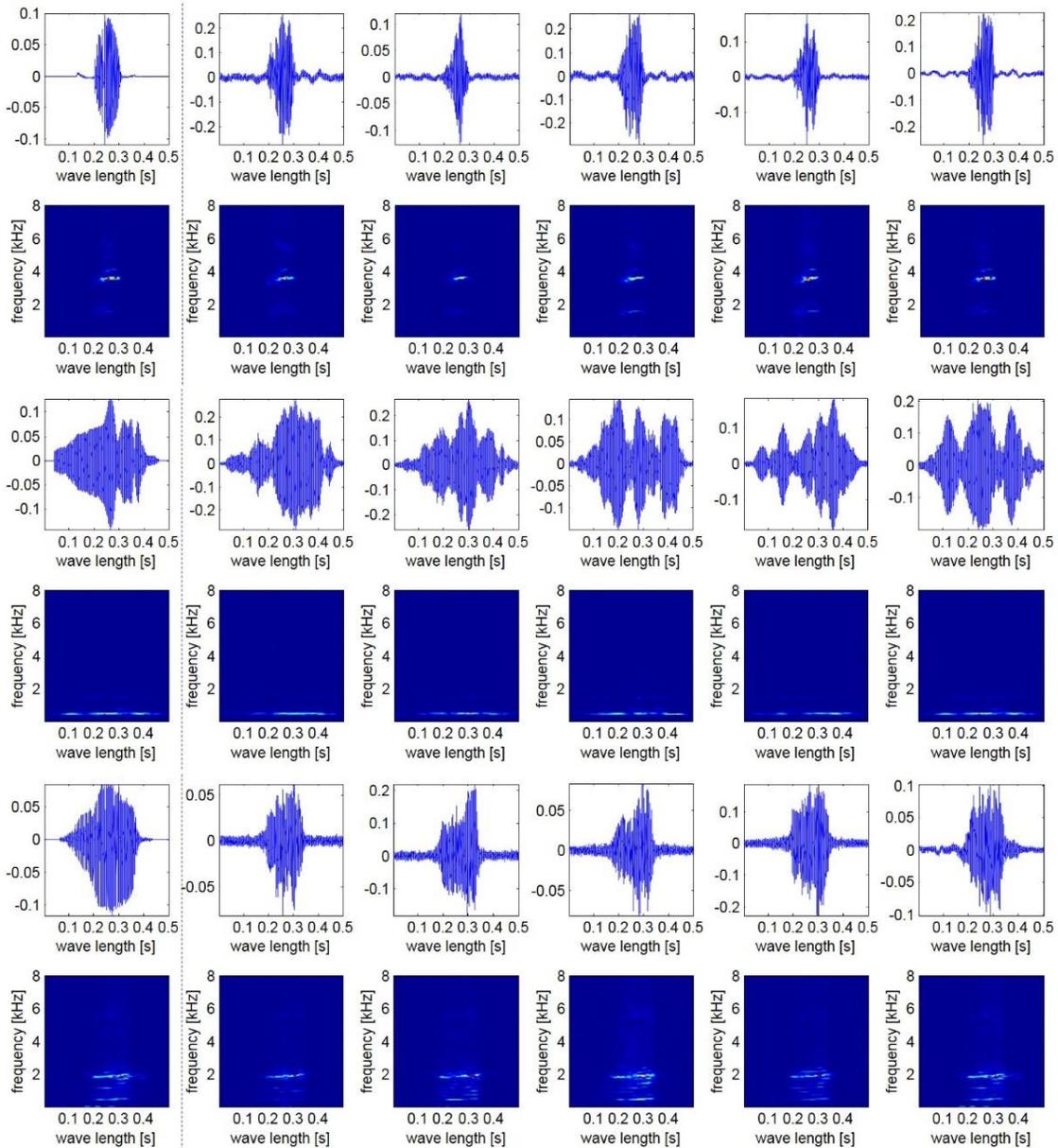

Fig. 6. Audio waveform and spectrogram of three species; The first two rows correspond to SuwFrog, middle two rows to RedFrog, and the last two rows to NarFrog. A real waveforms for each species are shown in the first column, and five synthesized waveforms are shown in the following columns.

*3) Quantitative comparison*

To quantitatively compare to other prominent methods, an inception score which is typically used to measure the quality of generated data based on adversarial learning [28], [37], [43] is applied. In order to calculate the score, audio classifier based on CNN depicted in Fig. 7 was trained for modeling a posterior probability instead of inception network. The real call sounds were transformed to spectrograms by applying STFT with the frame set to 256 samples with 75% overlap with the previous frame, and with a hamming window. After the spectrogram is resized to 128x128, it is inputted to the CNN. In all convolution layers, the filter size is set to 3x3. BN is applied to the results of the convolution, and ReLU is used as the activation function in the convolution and fully-connected layers. The posterior probability required to calculate the inception score can be obtained in the output layer of the CNN.

Table II shows the inception scores according to methods. Note that the proposed architecture depicted in Fig. 1 is applied to all other GAN methods without regularization, and the case of using proposed objective function composed by $f_{WGAN}$ and regularization term is denoted *Proposed* in Table II. Additionally, WaveNet [36], which is a generative model based on maximum likelihood not adversarial learning is also compared because it is well-known generative model for synthesizing audio waveforms.

It is known that GAN suffers from unstable learning, and GAN has the lowest score among considered cases. The second worst score is that of WaveNet. Although the WaveNet generates realistic waveform for NarFrog call sound, however it fails to generate other frog sounds. As mentioned previously, a large-sized receptive field has a problem to generate short time waveform like amphibian sounds. Note that the size of receptive fields in WaveNet is set to about 300 ms [36], however the average length of frog call sounds is about 150 ms. (as an example, Fig. 6 shows that the length of call sounds for SuwFrog is about 100ms.) From the results, the proposed method based on WGAN attains the greatest score although LSGAN and WGAN seem to improve the quality of the generated data. Note that the training termination criterion is met if the difference in generator loss between epochs falls to within 1.0e-6. In case the loss is never bounded, training stops at 10000, the maximum number of epochs allowed.

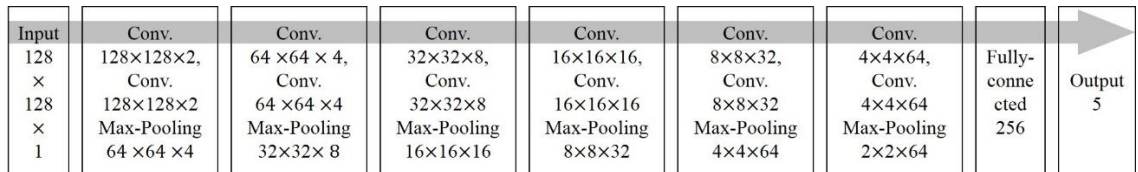

Fig. 7. CNN architecture applied to calculate inception score and perform an amphibian call classification

TABLE II
INCEPTION SCORES OF OTHER METHODS FOR RAW AUDIO WAVEFORM GENERATION

| GAN | LSGAN | WGAN | WGAN_LP | Proposed | WaveNet |
|---|---|---|---|---|---|
| 1.1483 | 2.3645 | 3.9795 | 2.3597 | **5.4136** | 2.0257 |

## V. EXPERIMENT II. AMPHIBIAN SOUND CLASSIFICATION

As mentioned previously, the idea of generative models for augmenting data is beginning to attract attention among researchers because while the data required for deep learning (particularly labeled data) is expensive, the technique has had great success in many tasks. This section shows the effectiveness of using the synthetic data produced by the proposed network to train a deep classifier.

### A. Database

For the target classes except RedFrog, the two datasets, *RealTrn* and *RealTst*, are determined by randomly and exclusively selecting 150 and 200 audio streams, respectively, from each class dataset in TABLE I. In the case of RedFrog, 150 audio streams are randomly selected for *RealTrn*, and all the data is assigned to *RealTst* due to a lack of data. The classification performance is summarized by the average of 5-fold cross-validation test results.

### B. CNN in amphibian sound classification

As CNN has been shown to perform well in animal sound classification [44], [45], CNN which is considered for calculating inception score is applied for this task. In the same manner, the resized spectrograms is inputted to the CNN. And, predicting label was performed in the output layer based on softmax criteria.

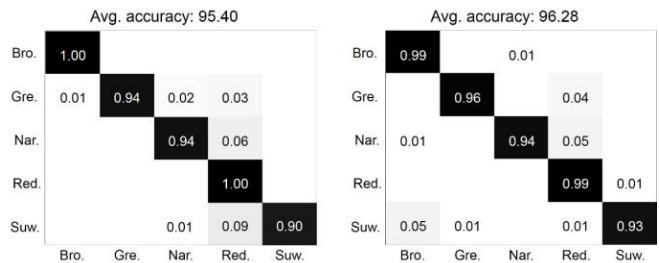

Fig. 8. Confusion matrices of the results of Experiment 1 (left) and 2 (right)

TABLE III
THE RESULTS OF AMPHIBIAN CALL CLASSIFICATION USING CNN TRAINED BY SYNTHETIC DATASET

| # of data / class | Classification rate in each class [%] | | | | | Avg. |
|---|---|---|---|---|---|---|
| | Bro. | Gre. | Nar. | Red. | Suw. | |
| 75 | 23.50 | 91.00 | 85.50 | 97.59 | 82.00 | 75.72 |
| 150 | 19.00 | 91.50 | 90.50 | 98.80 | 78.00 | 75.56 |
| 225 | 43.50 | 91.50 | 91.50 | 98.19 | 91.00 | 83.14 |
| 300 | 80.50 | 92.50 | 90.50 | 99.40 | 92.50 | 91.08 |
| 275 | 95.00 | 95.00 | 92.50 | 98.80 | 91.00 | 94.46 |
| 450 | 93.50 | 95.50 | 92.00 | 98.80 | 92.50 | 94.46 |
| 525 | 92.00 | 96.50 | 93.50 | 98.80 | 90.50 | 94.26 |
| 600 | 99.00 | 96.50 | 93.50 | 99.40 | 93.00 | 96.28 |

### C. Classification results

#### 1) Experiment 1

This experiment has been performed to evaluate the CNN architecture for amphibian call classification. The *RealTrn* sets of all classes were used to train the CNN and the *RealTst* sets of all classes used to assess the performance. The class average classification rate is 95.40% as shown in Fig. 8. Thus, according to the results, the amphibian call classification can be appropriately performed using the CNN depicted in Fig. 7.

#### 2) Experiment 2

In this experiment, the CNN is trained with synthetic streams, and evaluated with the *RealTst* sets used in Experiment 1. In this case, the performance is nearly the same as in Experiment 1, although only the synthetic data was used for training. It is apparent that the distribution of the corresponding synthetic data matched that of the real data. TABLE III shows the results according to the quantity of synthetic data. As expected, the more training data is used, the better class average classification rate is achieved.

#### 3) Experiment 3

In this experiment, we demonstrate the effectiveness of data augmentation by the proposed network in a classification task. The CNN is trained using both the *RealTrn sets* and the synthetic data used in Experiment 2, and then evaluated using the *RealTst* sets used in previous two experiments. The performance assessments are conducted according to the quantity of synthetic data. The results are summarized in TABLE IV that also shows the composition of the training data in the left column. From the table, the best average classification rate represents an improvement of about 1.1% over the results of Experiment 1. Additionally, the CNN is quickly converged when synthetic data is used in training as shown in Fig. 9. When the quantity of training data is augmented by synthetic data, the processing time per epoch takes longer than the case of only real data. However, the number of epochs required for convergence is significantly reduced by adding synthetic data. Note that the numbers with a single bracket in Fig. 9 indicate the conditions presented in TABLE IV.

TABLE IV
THE RESULTS OF AMPHIBIAN CALL CLASSIFICATION USING CNN TRAINED BY BOTH REAL AND SYNTHETIC DATASETS

| # of data / class (real+syn.) | Classification rate in each class [%] | | | | | Avg. |
|---|---|---|---|---|---|---|
| | Bro. | Gre. | Nar. | Red. | Suw. | |
| Real only | 100.00 | 93.50 | 93.50 | 100.00 | 90.00 | 95.40 |
| 1) 150+75 | 100.00 | 94.50 | 94.50 | 100.00 | 89.50 | 95.70 |
| 2) 150+150 | 100.00 | 94.50 | 94.00 | 100.00 | 92.00 | 96.10 |
| 3) 150+225 | 100.00 | 96.00 | 94.00 | 100.00 | 86.50 | 95.30 |
| 4) 150+300 | 100.00 | 97.00 | 94.00 | 100.00 | 86.50 | 95.30 |
| 5) 150+375 | 100.00 | 94.50 | 94.00 | 100.00 | 83.50 | 94.40 |
| 6) 150+450 | 100.00 | 96.00 | 94.00 | 100.00 | 87.50 | 95.50 |
| 7) 150+525 | 100.00 | 94.50 | 93.50 | 100.00 | 87.00 | 95.00 |
| 8) 150+600 | 100.00 | 97.00 | 95.50 | 100.00 | 90.00 | 96.50 |

## VI. CONCLUSIONS

This paper introduced a signal generation network based on adversarial learning, and demonstrated the effectiveness of data augmentation in amphibian sound classification. The generator and discriminator for adversarial learning were respectively composed by MLP and CNN, and they were trained by maximizing Wasserstein distance. Once converged, signal waveforms can be produced by the generator. The reproducibility and productivity of the proposed network were verified experimentally by using mathematically produced sinusoidal waveforms and amphibian call sounds that were manually collected in the natural habitats of each species. Graphical comparisons of time-domain waveforms and spectrograms and quantitative comparisons using the inception score clearly showed that the synthetic data closely resembles the target signal. Overall, it was demonstrated that the proposed approach of data augmentation by direct generation of synthetic audio streams improved the CNN based classification rate and its training efficiency when both the real and the synthetic data were used to train the classifier. These results demonstrate that the proposed network generates an arbitrary signal that is composed of sinusoidal waveforms and can be used for training a deep network.

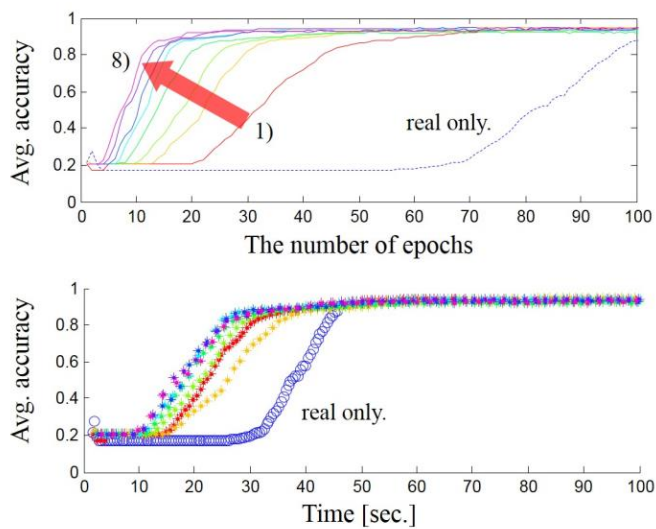

Fig. 9. Comparison of training speeds according to the composition of training data

## VII. ACKNOWLEDGEMENT

Authors of Korea University were supported by Korea Environment Industry & Technology Institute (KEITI) through Public Technology Program based on Environmental Policy, funded by Korea Ministry of Environment (MOE) (2017000210001). David Han's contribution was supported by the US Army Research Laboratory.

**Sangwook Park** received his B. S. degree in electrical and electronics engineering form Chung-Ang University, Seoul, South Korea, in Feb. 2012. Since 2012, he has been in the integrated MS. and Ph D. program in electrical engineering at Korea University, Seoul, South Korea. In Aug. 2017, he received a Ph D. degree from Korea University. From Sep. 2017 to Aug. 2018, he was a research professor enrolled in the BK21plus program of the school of electrical engineering, Korea University. Now, he is currently working in the Laboratory of Computational Audio Perception as a postdoc fellow in the Johns Hopkins University.

His research interests include machine learning using deep learning for pattern recognition and generation. In detail, he is working on acoustic event/scene recognition, acoustic spatial understanding, and acoustic signal generation using deep generative models.

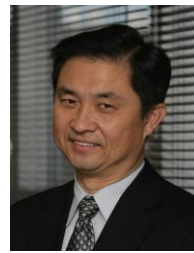

**David K. Han** (M'07-SM'13) received his B.S. degree from Carnegie Mellon University, and MSE and PhD from Johns Hopkins University. After years of serving as scientist at NSWC and ONR, he joined the University of Maryland in College Park in 2005 as a visiting professor. From

2007 to 2009, he was the Distinguished IWS chair professor in the Systems Engineering Department, US Naval Academy, Annapolis. In 2009, he returned to ONR as a program officer in Ocean Engineering and in 2012, as the Deputy Director of Research of ONR.

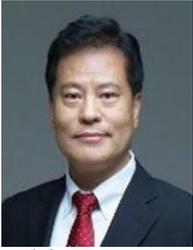

**Hanseok Ko** (M'95-SM'00) received his B.S. degree from Carnegie Mellon University, in 1982, M.S. degree from Johns Hopkins University, in 1988, and Ph.D. degree from the CUA, in 1992, all in electrical engineering. At the onset of his career, he was with the WOL, Maryland, where his work involved signal and image processing. In 1995, he joined the faculty of the Department of Electronics and Computer Engineering at Korea University, where he is currently Professor. His professional interests include speech/image signal processing for pattern recognition, multi-modal analysis, and intelligent data fusion.